\documentclass{article}
\usepackage[T1]{fontenc} 
\usepackage[utf8]{inputenc} 
\usepackage{ismir,amsmath,cite}
\usepackage[hyphens]{url}
\usepackage{kotex}
\usepackage{graphicx}
\usepackage{color}
\usepackage{tabularx}
\usepackage{musicography}
\usepackage{blindtext}
\usepackage{lineno}

\title{Finding Tori: Self-supervised Learning for Analyzing Korean Folk Song}

\oneauthor
  {Danbinaerin Han$^1$ \hspace{2cm} Rafael Caro Repetto$^2$\hspace{2cm} Dasaem Jeong$^1$} 
  {$^1$Department of Art \& Technology, Sogang University \hspace{1cm} $^2$Kunstuniversität  Graz \\ {\tt\small naerin71@sogang.ac.kr, rafael.caro-repetto@kug.ac.at, dasaemj@sogang.ac.kr}}

\def\authorname{D. Han, R. Caro Repetto, and D. Jeong}

\begin{document}

\maketitle
\begin{abstract}
In this paper, we introduce a computational analysis of the field recording dataset of approximately 700 hours of Korean folk songs, which were recorded around 1980-90s. Because most of the songs were sung by non-expert musicians without accompaniment, the dataset provides several challenges. To address this challenge, we utilized self-supervised learning with convolutional neural network based on pitch contour, then analyzed how the musical concept of \textit{tori}, a classification system defined by a specific scale, ornamental notes, and an idiomatic melodic contour, is captured by the model. The experimental result shows that our approach can better capture the characteristics of tori compared to traditional pitch histograms.
 Using our approaches, we have examined how musical discussions proposed in existing academia manifest in the actual field recordings of Korean folk songs.

\end{abstract}
\section{Introduction}\label{sec:introduction}
Folk songs are considered as a musical language. Not only that, but they are also regarded as the root of all traditional music. Folk songs are potent embodiments of a region's cultural and linguistic characteristics, serving as a foundation for all artistic and musical developments since their inception. It is believed that research for folk music can provide new inspiration for existing art music research, and facilitate an interdisciplinary approach that encompasses both music and other fields.

In Korean musicology, there have been ongoing discussions aimed at identifying regional characteristics in folk music. Through numerous debates, \textit{tori} is used as the most representative theory in many studies. 
Even though the use of tori as a term to describe the musical characteristics of Korean folk music is widespread and its utility is well acknowledged, the existing tori classification methods were unable to fully explain the features of the actual music. Various scholars have defined the different tori according to musical characteristics such as scales (intervals between notes), primary notes, ornamentation, ending note and idiomatic melodic patterns in folk songs, leading to a refinement process for the classification of tori. So far, there are still ongoing opinions, controversies, and discussions concerning the existing tori\cite{shin2016, sung2020koreamusic}.

In order to conduct valuable discussions on analyzing folk songs, the relationships within the audio must be examined and systematically shared. However, checking numerous audio files individually is extremely time-consuming and difficult, so previous research has focused on analyzing small amounts of audio. 
The most common method used by musicologists has been transcription, which involves listening to the music and notating it in a specific music notation system. However, the task of transcribing orally transmitted music into a music notation system has inherent limitations\cite{kim2013transciption}, such as quantizing the pitch and rhythmic features of folk songs. 

For these reasons, there have been several research on analyzing folk music empowered with methods derived from music information retrieval (MIR) on audio recording. 
For example, Indian \textit{raga} music has been analyzed based on first-order pitch distrubutions \cite{koduri2012raga}, and for traditional three-part Georgian singing, F0-based tonal analysis\cite{rosenzweig2019detecting} or development of tuning systems\cite{scherbaum2022tuning} has been introduced. \cite{Kroher2018} also presented analysis using audio-signal processing on Flamenco and Arab-Andalusian vocal music.


In this paper, we take a computational approach to analyze a vast corpus of Korean folk songs, utilizing deep-learning-based methodologies. Our primary aim is to investigate the connection between conventional musicological classifications of Korean folk songs, \textit{tori}, and the actual field recordings. Through a comparison of our analytical results with established musicological frameworks, our objective is to offer insights on identifying meaningful clustering or distinguishable features among Korean folk songs. We also share our code and metadata that contains links to original audio with our manual tori labels for 218 songs\footnote{\url{https://github.com/danbinaerinHan/finding-tori}}.

\section{Korean Folk Songs and Tori}

\subsection{Dataset}\label{subsec:dataset}
The ``Anthology of Korean Traditional Folksongs'' is an audio collection consisting solely of traditional Korean folk songs \cite{choi2000}. As part of a cultural project conducted by Munhwa Broadcasting Corporation (MBC) from 1989 to 1995, the folk song collection was compiled under the direction of Sang-il Choi. The available audio contains 15,861 songs, with approximately 700 hours in total length. 

The audio was field-recorded across 153 city/county and 1,010 villages in South Korea. 
The metadata accompanying the recordings includes title, machine-readable lyrics, regions, recording dates, and the singer's name and age. The region represents the administrative districts of South Korea at the time of recording, which consists of nine categories in total. 
Due to its extensive audio corpus and detailed accompanying metadata, this dataset has become a crucial resource for research on Korean folk music within the domestic academic community\cite{bae_ingyo2003,Choi2004,kim1999}.
We obtained the audio material through crawling the original website,\footnote{\url{http://urisori.co.kr/urisori-en/doku.php/}} which has been hosted by MBC since April 2022, where every audio and metadata is openly published. We received official approval for using these data for research purposes from MBC.

\subsection{Tori}\label{subsec:tori}
Tori was proposed by a Korean musicologist named Bo-hyeong Lee in the early 1980s to explain the musical characteristics of regional folk songs\cite{lee1982gyungseo, lee1983tori}. 
Prior to the development of tori theory, Korean music academia used the terms \textit{jo} and \textit{cheong} to explain the musical features of folk songs, based on the tonal center and intervals between notes\cite{baek2004}. 
However, Bo-hyeong Lee argued that \textit{jo} and \textit{cheong} are insufficient in capturing the unique musical characteristics of Korean folk music. Therefore, he proposed the tori classification system as an alternative approach, which covers primary notes, ornamentation, ending note and idiomatic melodic patterns.


The most widely-used tori classification method divides songs into four categories. We encourage readers to refer Figure \ref{fig:pitch_histogram}, which presents scale of three tori with staff notation. The pitch name in this section is used as an conventional representation, not absolute pitches.
\textbf{Gyung-tori} utilizes five notes and is mainly sung in the capital region. It often uses a gentle vibrato overall and finessed ornamental melody. 
\textbf{Menari-tori} is widely distributed throughout the eastern regions and the entire Korean Peninsula. When the melody ascends, it leaps through notes, while during the descending melody, it comes down through passing notes.
\textbf{Yukjabaegi-tori} is commonly found in the southern regions of the Korean Peninsula. It is characterized by a thick and vibrant vibrato in G note and passing briefly through E\musFlat{} in a descending melody from E\musFlat{} to D. 
\textbf{Sushimga-tori} commonly appears in the northwestern region of the Korean Peninsula, characterized by its unique vibrato inflecting upward.

\subsection{Tori annotation}
As the dataset did not include any tori labels, one of the authors, who has over 10 years of experience in Korean traditional music, selected a subset of 218 high-quality recordings and manually annotated them in terms of tori classification.

To create this tori-subset, we focused on identifying clear musical characteristics present in the tori classification method, such as pitch scale, ornamentation, and idiomatic melodies, rather than considering the audio's recorded region. We found that there are not many instances of sushimga-tori in the audio dataset, as the dataset predominantly consists of folk songs from the central and southern regions of the Korean Peninsula. Also, we observed that songs belonging to menari-tori appeared not only in the eastern region but also nationwide. In conclusion, we mainly focused on the remaining three types of tori, \textit{gyung-tori}, \textit{yukjabaegi-tori}, and \textit{menari-tori}. Finally, audio recordings that were closer to speech or chanting (non-musical) and did not exclusively belong to any other subdivided tori types were labeled as `others'. We have labeled totally 218 songs: 65 gyung, 73 menari, 49 yukja, and 31 others.

\section{Methodology}\label{sec:method}
In this study, we take an approach of representing the musical characteristics of a given folk song using a high-dimensional single vector, which can also be called an embedding of the song. If this embedding accurately represents the musical characteristics of tori, it can be utilized for various purposes including tori classification, song similarity searches and clustering songs in the corpus based on tori similarity.

\begin{figure}[t]
 \includegraphics[width=0.95\columnwidth]{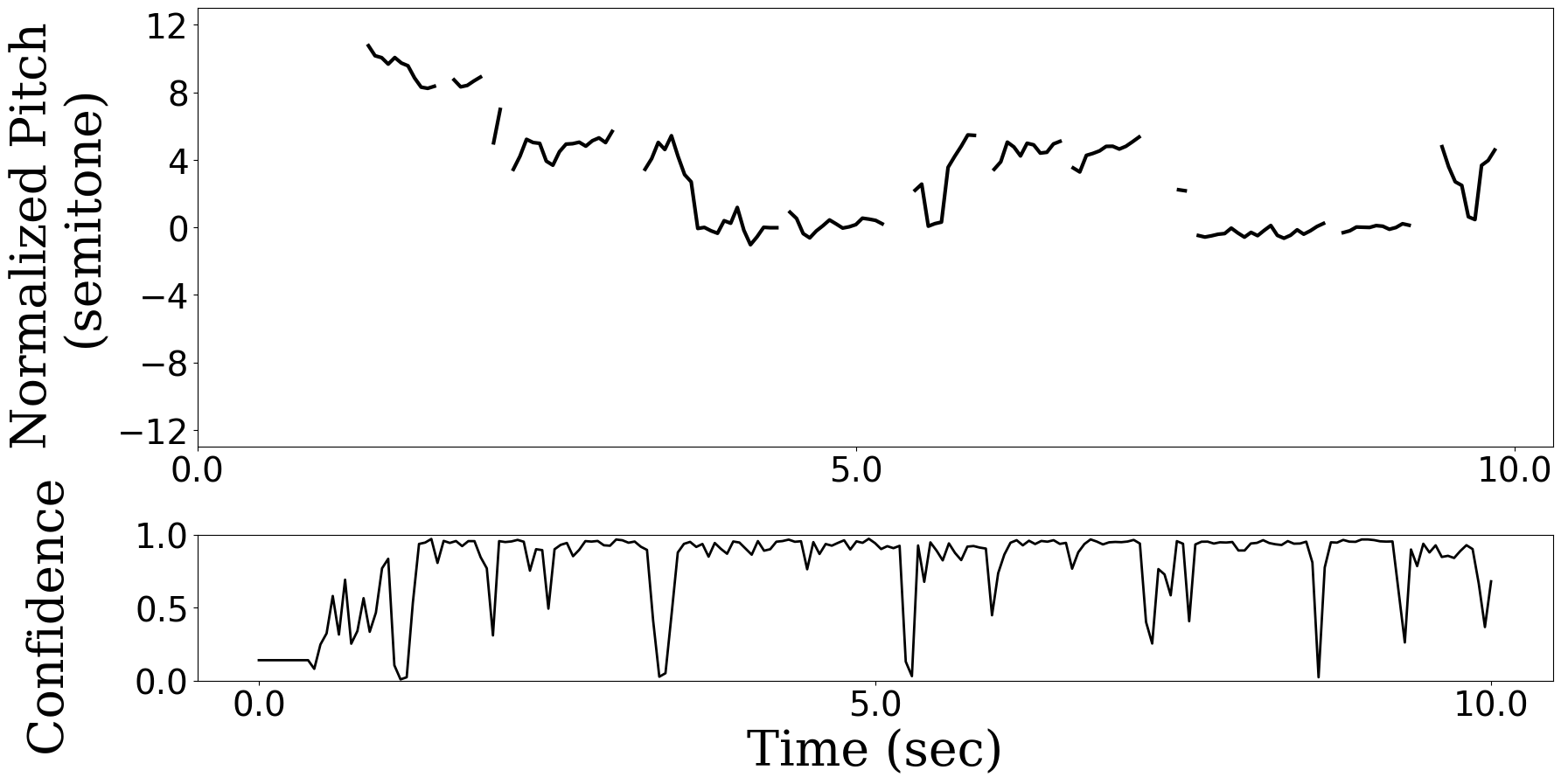}
 \caption{Example of pitch contour from the dataset extracted by CREPE. F0 value under confidence of 0.8 was masked out}
 \label{fig:contour}
\end{figure}

Even though the dataset is provided with audio recordings, we focus on the contour of fundamental frequency (F0) rather than using audio directly. By doing so, our embedding could better capture the melodic features of the songs rather than timbral characterstics such as the dialect of the singer.
To extract the F0 contour from each audio recording, we utilized CREPE \cite{kim2018crepe}, a widely-used CNN-based model. It extracts F0 value for every 10 ms as well as confidence score ranging from 0 to 1. Figure \ref{fig:contour} illustrates an example of an extracted pitch contour.

\subsection{Pitch Histogram}
Since the characteristics of a musical system can be largely studied from the pitch distribution of its melodies,
there has been numerous research using pitch histograms to analyze traditional musics from different culture, such as for Indian Carnatic music \cite{koduri2012characterization}, Turkish Makam \cite{gedik2010pitch}, Arab-Andalusian music \cite{pretto2018nawba}, or Iranian \textit{dastgāhi} music \cite{nikzatkdc}. Therefore, we applied a similar approach to Korean folk music.

One of the important features one has to know to exploit pitch histogram is the tonic of the song so that one can normalize the pitch histograms of each song into relative pitch, rather than directly using absolute pitch value.
During our preliminary experiment, we found that the tonic of each recording in the dataset is usually the most frequently appearing pitch in the song. Therefore, we determined the tonic by counting the number of appearances of each pitch in terms of time frames, utilizing a 100 cents range to count pitch (i.e. MIDI pitch 60.49 and 59.51 are counted as the same pitch). As the center pitch may differ from the typical 440 Hz tuning, we identified the best matching pitch by adjusting the pitch center by increments of 10 cents and selecting the pitch that showed the maximum pitch count.

\subsection{CNN contour encoder}
While the pitch histogram can offer a comprehensible representation of each song, it cannot fully capture the concept of tori as it does not consider the relationship between each note in the scale. Additionally, because most of the recordings were sung by non-professional singers without instrumental accompaniment, the intervals of singing and tonic frequency has signifcant amount of noise. 
Therefore, we employ a convolutional neural network (CNN) to learn the representation of a given pitch contour. This approach allows us to better capture the underlying musical characteristics of each song in relation to tori classification.

The problem is setting the training objective of the CNN model. One option is to train the model in a supervised manner, as typically done in other MIR tasks, such as classification tasks with an annotated dataset. In our dataset, region labels, which represents the administrative region where the recording took place, can be used for training labels. It can be regarded as a reasonable approach considering that tori has strong correlation with the regional characteristics of each folk song. However, we discovered several different types of tori in a single recorded region, which would make it difficult to learn distinguishable musical characteristics only using the region label.

\begin{figure}[t]
 \includegraphics[width=0.95\columnwidth]{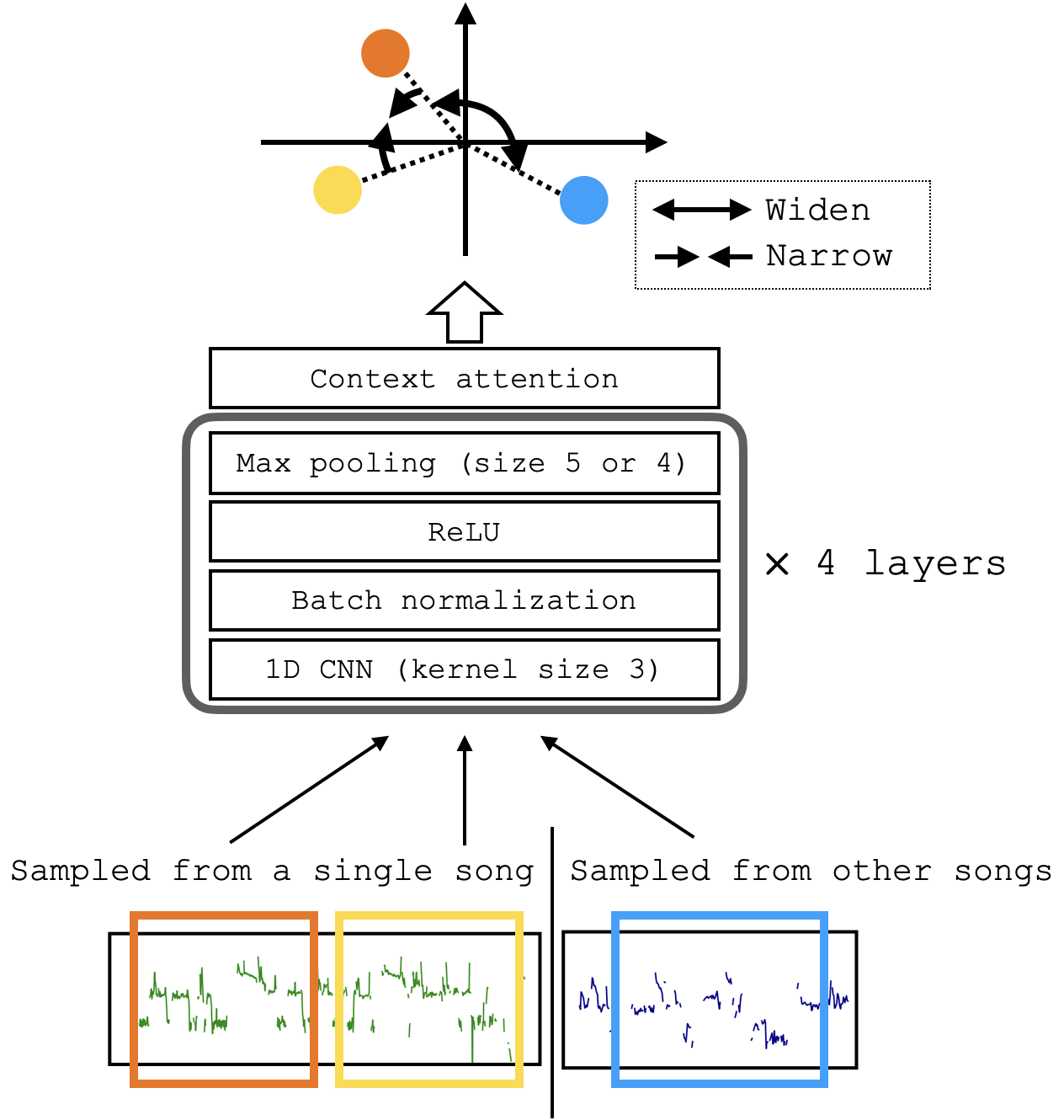}
 \caption{Self-supervised learning with triplet loss using cosine similarity, with four layers of convolutional neural network}
 \label{fig:SSL}
\end{figure}

To address this issue, we propose to adopt a self-supervised representation learning, so that we can obtain a representation of given pitch contour that is consistent within a song without extra annotated labels as shown in Fig. \ref{fig:SSL}. This approach of exploiting intra-song similarity has been widely used for music audio representation learning \cite{lee2019representation, spijkervet2021contrastive}. We use triplet loss with hinge margin as Equation \ref{eq:loss},
\begin{equation}
    L = \max(0, m - \text{Sim}(v_a, v_p) + \text{Sim}(v_a, v_n)) 
\label{eq:loss}
\end{equation}
where $m$ denotes hinge margin, $\text{Sim}(v_a, v_{p, n})$ represents cosine similarity between anchor vector $v_a$ and positive vector $v_p$ or negative vector $v_n$.

As presented in Fig. \ref{fig:SSL}, the model consists of 4 layers of 1D convolution layer, each with a kernel size of three and channel size of 64, 128, 256, and 256, respectively, each followed by batch normalization layer. We used a max pooling layer between each convolutional layer with size 5 for the first layer and 4 for the second and third layers. On top of the convolutional stack, we used context attention to summarize the arbitrary length of vectors, which was initially proposed in hierarchical document classification \cite{yang2016hierarchical}, with a slight modification of multi-head attention weight \cite{vaswani2017attention} instead of single-head. Context attention is a type of attention mechanism that uses an independent learnable parameter named context vector as a query while using the input vectors as key and value. The final embedding size of the model was 256. 

The input is a pitch contour in the shape of $T \times 2$, where $T$ denotes the number of time step of pitch contour. Throughout the experiment, we used the frame rate of 20 Hz, which means that 30 seconds of pitch contour is converted to 600 time steps. During the training, we randomly slice the contour to have 30 seconds length, while using the entire contour in the test or visualization. Two channels are tonic-normalized MIDI pitch and confidence of the F0 estimation at that time frame. Since the automatic F0 estimation includes noise, especially in the unsung silent part, we masked the estimated pitch value to zero if the confidence is lower than 0.8.

\section{Experiments}\label{method:experiments}

\subsection{Evaluation Design}
In the experiments, we explored the quantitative relation between tori and different embedding schemes to see which type of embedding can explain the concept of tori better. To evaluate the correlation between tori and the given embedding scheme, we exploit our tori-labeled subset with two metrics. 

The first metric is the ranking of cosine similarity. If the embedding shares important characteristics with the concept of tori, we can expect that songs with high cosine similarity in the given embedding space share the same tori with the query song. Therefore, we calculated the cosine similarity between each song in the tori-subset, and obtained normalized discounted cumulative gain (nDCG), which is a frequently used metric to evaluate the quality of search results. If all the other $n$ songs with the same tori are ranked within $n$-th order in the similarity, nDCG becomes one.

The second metric is the tori classification accuracy using a random forest classifier. If the embedding includes essential characteristics that define tori, one can expect that it can be exploited to classify the tori for a given song. Among many options, we used a simple random forest classifier. The random forest classifier with 100 trees was trained with 75\% of the tori-set and tested with the remaining 25\%. For each embedding scheme, we repeated the procedure 30 times with different train/test split and reported mean and deviation of accuracy to ensure that the results were not dependent on the specific dataset split.


\subsection{Training}
While a pitch histogram can be obtained from a given song without additional training data, CNN models require training procedures. We employ F0 contour extracted from the ``Anthology of Korean Traditional Folksongs'' to train our model.

Some of the songs in the dataset were recorded with multiple singers or percussion accompaniment. Because these can degrade the performance of F0 estimation, we excluded them while training the CNN model. Instead of manually filtering the dataset, we used a CNN-based sound event detection model \cite{kong2020panns} to calculate the activation of `choir' and `percussion,' which are included among the model's event vocabulary. We filtered the song by the maximum activation value exceeding a certain threshold, which was manually decided by observing the activation distribution across the dataset. 

The CNN model was trained for 25,000 updates. For the self-supervised learning using triplet loss, we used eight negative samples with one positive sample, and hinge loss using a margin of 0.4. For the region-supervised training, we use cross-entropy loss with class weight to address the class imbalance issue. The architecture of the CNN remained the same, but we added a fully connected layer to project the embedding to logits for nine different region categories. We used the Adam optimizer with an initial learning rate of 0.001 and batch size of 128. The entire training process was conducted on a single RTX A5000, which took approximately 10 minutes to complete 25,000 updates. We validated the model's training procedure and hyperparameters using a specific split of the tori-subset and chose the simplest setting to avoid overfitting the hyperparameters to the small subset.

\subsection{Results}

\begin{table}[]
\resizebox{\columnwidth}{!}{
\begin{tabular}{l|c|c}
                        & \multicolumn{1}{c|}{Similarity Ranking}   & \multicolumn{1}{c}{Random Forest Classifier}  \\[1.2ex]  
Embedding &
  \multicolumn{1}{c|}{nDCG} & 
  \multicolumn{1}{c}{Accuracy}  \\[1.2ex] \hline 
Hist. (25 bin)    & 0.783                        & 0.744 $\pm$ 0.058            \\[1.2ex]
Hist. (124 bin)   & 0.777                        & 0.722 $\pm$ 0.054            \\[1.2ex]
CNN (region ) & 0.792                          & 0.634 $\pm$ 0.055            \\[1.2ex]
CNN (SSL)   & \textbf{0.853}                  & \textbf{0.848} $\pm$ 0.039        
\end{tabular}%
}
\caption{Experiment results on the tori subset. Hist. denotes normalized pitch histogram, specified with the number of bins to cover two-octave range. region and SSL denotes the region-supervised and self-supervised learning.}
\label{tab:results}
\end{table}

The evaluation result of each embedding scheme is presented in Table \ref{tab:results}. As tori is closely related to the use of pitch, the pitch histogram showed about 74 \% accuracy when combined with a random forest classifier, and 0.783 or 0.777 nDCG in the similar song search. The result shows that increasing the resolution of the histogram does not achieve better performance. The performance of CNN embeddings trained with region classification was worse than the pitch histogram in classification accuracy. However, the CNN embedding trained with self-supervised learning showed the best nDCG and classification accuracy, even though both CNN shares the same architecture except the final projection layer. 

The performance gain compared to the pitch histogram can be caused by the fact the concept of tori includes not only the pitch scale the song uses but also ornamentation and idiomatic phrases it, which can be easily captured by contour CNN.

The result also shows that the region label did not help to learn tori-related characteristics. Even though the concept of tori is strongly related to each region, folk songs were widely spread nationwide at the time of the recording, which made it difficult to learn coherent musical characteristics from the region label.

It is also worth noting that self-supervised learning did not use any other musical knowledge that is related to tori, except that it was trained to extract constant embedding throughout a given song regardless of the segment position. From a machine learning point of view, the model has to extract an embedding that can explain the entire song or distinguish it from other songs from a given fragment. The high correlation between the trained embedding and tori labels shown in the evaluation implies that the concept of tori is clearly related to the distinguishable characteristics of each Korean folk song.




\section{Analysis and Discussion}\label{sec:typeset_text}
\begin{figure*}[t!]
 \centerline{
 \includegraphics[width=1.3\columnwidth]{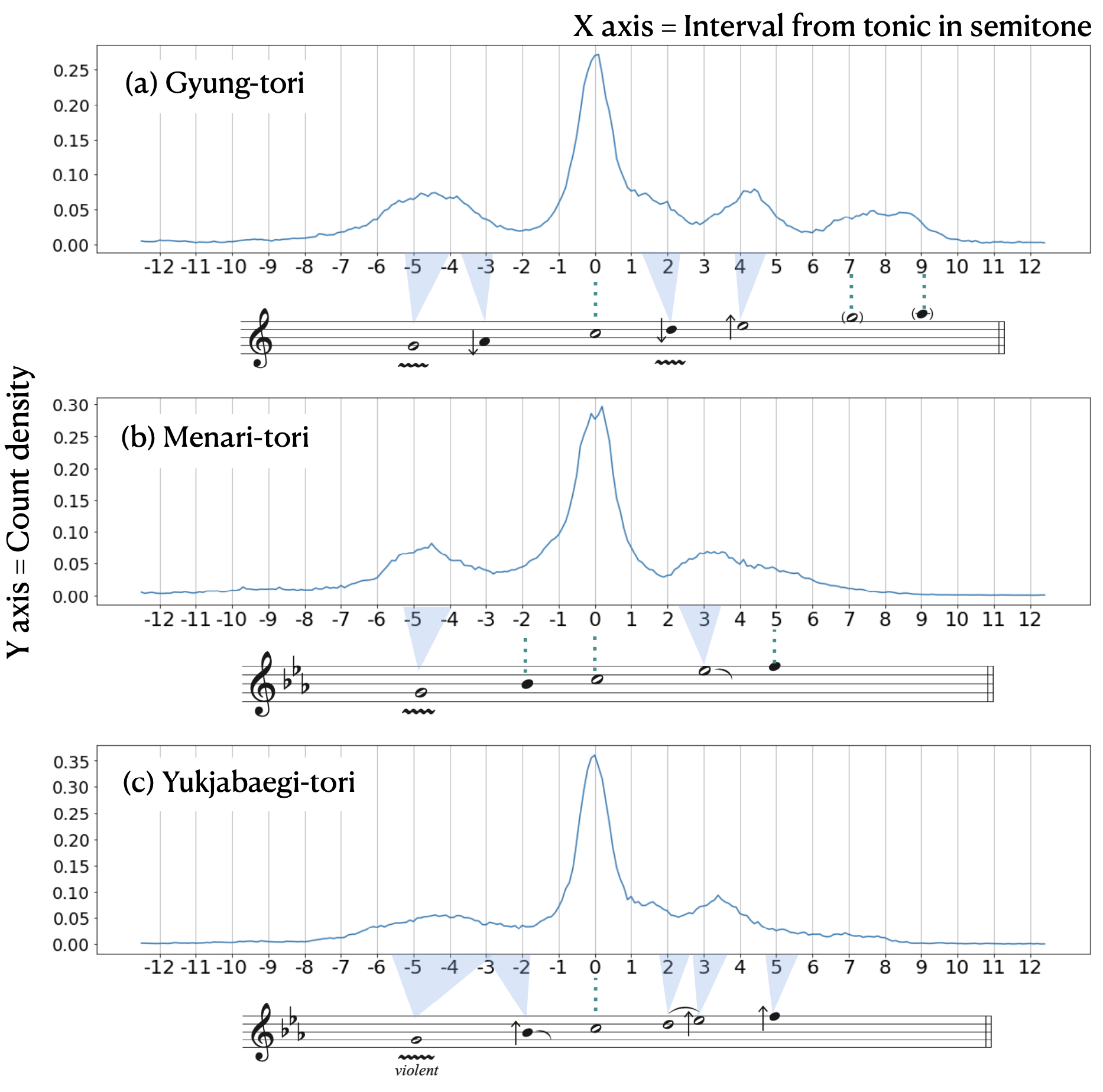}}
 \caption{Average pitch histogram of each tori obtained from the tori-subset and corresponding scale description.}
 \label{fig:pitch_histogram}
\end{figure*}
In this section, we introduce interesting musicological findings that we have found from the pitch histogram and embedding learned from self-supervised training.

\subsection{Pitch Histogram Analysis}\label{subsec:pitch_histogram}
To explain the scale and characteristics alongside with pitch histogram of each tori, we present the staff notation of each tori in Figure \ref{fig:pitch_histogram} based on various studies describing tori \cite{lee2012, kim2020gaeron, sung2020koreamusic}. 
In the score, the \textit{cheong} (a main note that generally appears most frequently) was fixed as the C5. The unfilled notes are primary tones, and the filled notes are passing tones. In cases where there is a vibrato mark below the note, it is called \textit{yoseong}, meaning the pitch generally oscillates up and down. In addition, the arrow on the left side of the note indicates that the pitch is slightly raised or lowered compared to the equal-tempered pitch. The slur mark flowing down to the right of the note represents \textit{twoeseong}, which means the pitch slides down.

While menari-tori and yukjabaegi-tori remain within a perfect fifth range from the center, in gyung-tori, there is a larger proportion of high notes (G5-A5) or more from the center. This indicates that the distribution of notes from low to high in gyung-tori is evenly spread. This could be due to the fact that folk songs of gyung-tori are more commonly found in popular folk songs than in those that appear nationwide.
Perfect 4th below the center tone is common in all Korean music. However for example in gyung-tori, there is vibrato in the G4 note, and the A4 note has a low pitch, resulting in a distinct distribution in that part. In menari-tori, it can be observed that the B\musFlat{}4 note is used as a passing tone between C5 and G4. Yukjabaegi-tori is characterized by a vibrant vibrato in the G4, resulting in a gentle distribution in the nearby pitch than other tori.

Whether a major third or a minor third is used in the scale is an important factor in distinguishing the regional musical characteristics of the eastern and western parts of the Korean Peninsula in the traditional method. Subtle pitch differences can be observed in all three tori. In gyung-tori, the D5 note is clearly distributed lower and the E5 note is definitely higher than the line appearing the equal-tempered tune. In yukjabaegi-tori, it is also confirmed that the E\musFlat{} note is slightly higher than the minor third note's 

Through the above analysis, we found that pitch histograms can be utilized to roughly identify musical characteristics, such as the range, interval from tones and subtle pitches, etc. However, we also acknowledged that this approach solely exhibits the constituent pitches and their frequencies, without accounting for other aspects like uses of melody and ornamentation and more.

\begin{figure*}[t]
 \centerline{
 \includegraphics[width=1.8\columnwidth]{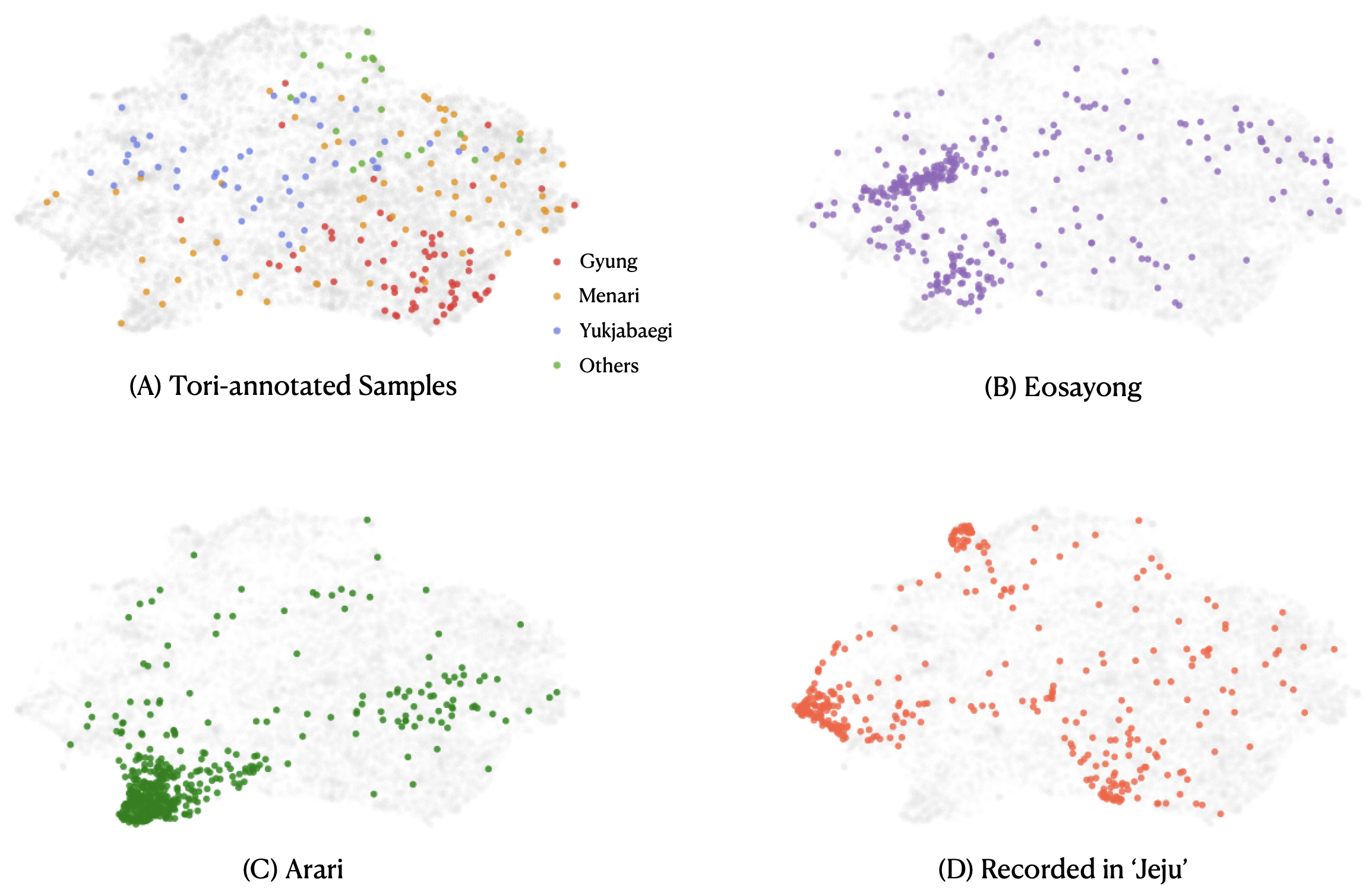}}
 \caption{UMAP visualization of folk song embeddings obtained from our self-supervised-trained CNN model. Note that no tori label was used to train the model.}
 \label{fig:umap}
\end{figure*}

\subsection{UMAP visualization}\label{subsec:umap_visualization}

We visualized the embeddings learned through self-supervised learning in 2D using UMAP, a dimension reduction technique that is frequently used for visualization of high-dimensional vectors in Figure \ref{fig:umap}. In the first figure, we demonstrated the distribution of four different tori labels. In addition, by examining the metadata information of the embeddings, we figured out several interesting topics from this visualization. We also implemented a web demo using this visualization where one can directly access the corresponding audio recording\footnote{\url{https://danbinaerinhan.github.io/korean-folksong-visualization/}}.

Menari-tori has a much wider distribution area and encompasses a broader musical range. In Figure \ref{fig:umap} (A), the menari-tori distribution can be seen to be wider compared to the other tori. We mentioned that several subdivided tori of it have been identified by scholars. Among them, eosayong-tori is a representative example of refining the musical characteristics of eastern folk songs\cite{kim1999}.
Eosayong-tori, a lamenting song sung by lumberjacks, was suggested to have distinct musical characteristics compared to menari-tori. For instance, in eosayong-tori, the lowest pitch is a semitone higher, and it concludes on the lowest pitch of the scale instead of the final pitch and the middle pitch found in menari-tori.
We examined embeddings corresponding to eosayong-tori in the metadata's title information, and they were clearly gathered in a different space than the area annotated as menari-tori, as presented in Figure \ref{fig:umap} (B).

Similar results can be observed in the following cases with \textit{arari}. \textit{Arirang}, having repetitive refrain lyrics such as ``arirang'' or ``arari'', is the most representative Korean folk song, with countless versions appearing in various regions. Among them, \textit{arari}, another name for \textit{Jeongseon arirang}, was sung primarily in Gangwon province before spreading nationwide in the 1930s\cite{Gang2008Arari, kim2011arirang}. Arari is also regarded as a separated tori from menari-tori for some researchers\cite{kim2017dongbu}, with its own slower tempo, monotonous skeleton tones, and the use of decorative tones. In Figure \ref{fig:umap} (C), we can see the clustering of entities with the title arari extracted from the metadata.

Another interesting example is songs from Jeju. Folk songs sung from Jeju island, which is the largest island in Korea, have fewer research results compared to other regions. In Jeju folk songs, the interval between the pitches is narrower compared to other provinces'. Due to these reasons, the traditional method of transcribing into staff notation was unsuitable for capturing the characteristics of Jeju music \cite{jo1997jeju, Lee2018jeju, shin2019jeju}. However, using our embeddings, we can identify three clusters as in Figure \ref{fig:umap} (D). It implies the potential usefulness of our approach to uncover the unique characteristics of folk songs in Jeju Island.


We also checked the UMAP of the pitch histograms but could not find clear clustering as in the presented examples.

\section{Conclusions}\label{sec:conclusion}
In this paper, we have presented our computational approach to obtaining a high-dimensional embedding vector for a given pitch contour, which allows us to analyze a vast amount of Korean folk songs.
Using these embeddings and a manually labeled subset, we have examined how musical discussions proposed in existing academia are manifested in our dataset. 
As our results cover various music characteristics, the learned embeddings can be utilized as a monitoring aid when dealing with numerous undefined data to review and find the tori. 
Also, we have discussed the methods and possibilities for utilizing and interpreting experimental results in the future research. Through this material, we can review and refine our understanding of the concept of the tori, and provide easily accessible resources and utilize them as appropriate evidence. Furthermore, there is potential to clarify the musical characteristics of regions that are distinct from other regions, like Jeju, about which existing research has not been as active. Ultimately, we expect this research to be valuable in illuminating the relationships and transformations of folk songs over time.

\section{Acknowledgement}
We wish to express our deepest gratitude to Sang-il Choi and his team for their invaluable work on the ``Anthology of Korean Traditional Folksongs''. Their effort to collect more than ten thousand unique recordings from across Korea has not only significantly contributed to the preservation of the rich tapestry of Korean folk music but has also created a monumental resource for academic research. Without their extraordinary efforts, our work would not have been possible. We sincerely thank them for laying such a robust foundation and for enabling future researchers and us to build upon it. We also appreciate MBC for kindly allowing us to utilize the dataset for this research.

The web demo was implemented with the kind help of Dongmin Kim.



\bibliography{ISMIR2023_template}


%
%
%
%
%

\end{document}